\documentclass[prl,aps,superscriptaddress,nofootinbib,floatfix]{revtex4}

\usepackage{graphicx}

\begin{document}

\title{Excitation of confined modes on particle arrays}                                                                                                                                                                                      
                  
\author{X. M. Benda\~na\footnote{These authors have contributed equally to the work.}}                                                                                                                                                       
\email{xm.bendana@csic.es}                                                                                                                                                                                                                   
\homepage{http://nanophotonics.es}                                                                                                                                                                                                           
\affiliation{Instituto de Qu\'imica-F\'isica ``Rocasolano'' - CSIC, Serrano 119, 28006, Madrid, Spain}                                                                                                                                       
\author{G. Lozano$^*$}
\author{G. Pirrucio}
\affiliation{Centre for Nanophotonics, FOM Institute AMOLF c/o Philips Research Laboratories, High Tech Campus 4, 5656 AE Eindhoven, The Netherlands}
\author{J. G\'omez Rivas}
\affiliation{Centre for Nanophotonics, FOM Institute AMOLF c/o Philips Research Laboratories, High Tech Campus 4, 5656 AE Eindhoven, The Netherlands}
\affiliation{COBRA Research Institute, Eindhoven University of Technology, P.O. Box 513, 5600 MB Eindhoven, The Netherlands}
\author{F. J. Garc\'ia de Abajo}
\affiliation{Instituto de Qu\'imica-F\'isica ``Rocasolano'' - CSIC, Serrano 119, 28006, Madrid, Spain}
\affiliation{Optoelectronics Research Centre, University of Southampton, Southampton SO17 1BJ, UK}

\begin{abstract}
We describe both theoretically and experimentally the existence and excitation of confined modes in planar arrays of gold nanodisks.
Ordered 2D lattices of monodispersive nanoparticles are manufactured, embedded in a silica matrix, and exposed to evanescent prism-coupling illumination, leading to dark features in the reflectivity, which signal the presence of confined modes guided along the arrays.
We find remarkable agreement between theory and experiment in the frequency-momentum dispersion of the resonances. Direct excitation of these modes reveals long propagation distances and deep extinction features.
This characterization of guided modes shows the great potential of metallic particle arrays for optical signal processing and distant sensing applications.
\end{abstract}

\maketitle

Metal nanoparticles are capable of producing high levels of confinement and enhancement of incident light \cite{LSB03}.
These appealing properties make them useful for sensing (e.g., through detection of plasmon-resonance shifts \cite{MV03,DTH06,KE06,SAT08,AHL08} and surface-enhanced Raman scattering \cite{NE97,KWK97,JXK05,TJO05,paper125}), cancer therapy \cite{QPA08,HSB03}, drug delivery \cite{LSH11}, improved photovoltaics \cite{CP08,AP10}, and catalysis \cite{AMO01,K02}.
These applications have attracted a considerable degree of interest on the optical properties of nanoparticles, which strongly depend on size, morphology, and composition \cite{MBS02,KCZ03,paper112}.
The interaction between neighboring particles can also produce changes in their optical behavior, which are dramatic when small particle gaps are involved \cite{B1982,NOP04,paper075}.
Additionally, long-range interactions lead to collective modes in ordered  \cite{paper142,KHH10} and disordered \cite{OER11} particle arrangements, which have been proposed for waveguiding in both 1D  \cite{QLK98,MKA03} and 2D \cite{paper105,paper142} configurations.

In particular, lattice resonances in 2D ordered particle arrays act as an effective way of creating guided modes \cite{paper090} that are strongly dependent on the choice of materials, particle size and shape, and lattice parameters, leading to distinct TE and TM modes \cite{paper142,RSB12}.
This configures a versatile platform for applications to distant sensing and waveguiding, which deserve a detailed characterization of the arrays.
Direct far-field sampling of  quasi-bounded modes in 2D arrays through diffracted orders have been extensively reported \cite{AB08,KSG08,CSY08,VGG09-2,VGG09,ZO11}.
Additionally, theoretical descriptions are available \cite{paper164}, which become particularly intuitive when assuming dipole inter-particle interactions \cite{paper090,paper142}.
However, guided modes in metal particle arrays have not been yet directly sampled (i.e., sampled without the assistance of diffraction to couple to far-field radiation), and a detailed experimental study of their optical behavior is still missing.

\begin{figure}
\centering\includegraphics[width=130mm]{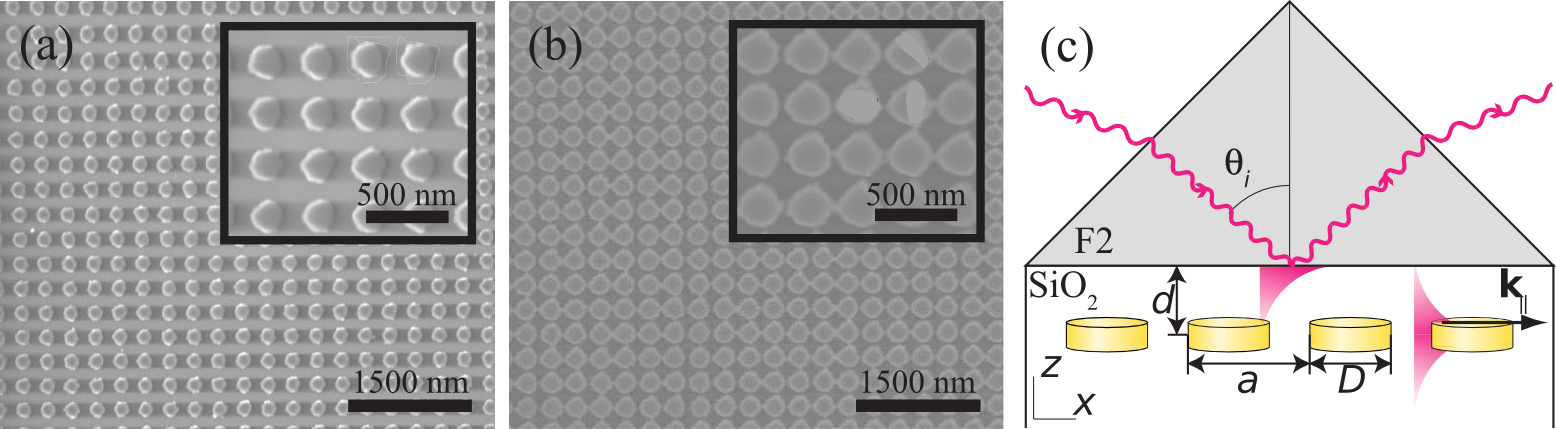}%
\caption{Images and scheme of the system under consideration. {\bf(a)} and {\bf(b)} SEM images of our nanoimprinted arrays.
The gold nanodisks are arranged in a square lattice with parameter $a=300\,{\rm nm}$. The height of the disks is $h=50\,{\rm nm}$ and the diameter is $D=200\,{\rm nm}$ in (a) and $D=250\,{\rm nm}$ in (b).
{\bf(c)} An array of golden disks is embedded in a homogeneous SiO$_2$ environment (refraction index $n_{{\rm SiO}_2}=1.46$) at a distance $d=600\,{\rm nm}$ from a prism interface.
This is the Otto configuration, whereby light undergoes total internal reflection inside the prism ($n_{{\rm F}_2}=1.67$) with incident angle $\theta_i$ and parallel wave vector $k_\parallel$ outside the light cone.
Light couples evanescently to the confined modes of the array, thus producing a measurable attenuated total reflection. \label{fig:scheme}}
\end{figure}

In this Letter, we report direct observation of guided modes in ordered gold-particle arrays through dips in their evanescent-field reflection using the  Otto configuration \cite{O1967,NH06}.
Gold nanodisks have been fabricated by imprint lithography \cite{VS07} and embedded in a symmetric silica environment [Fig.\ \ref{fig:scheme}(a,b)]. A high-index prism is then used for evanescent coupling of incident light [Fig.\ \ref{fig:scheme}(c)]. The reflected signal unambiguously shows the presence of guided modes.
We further report electromagnetic simulations for both the guided-mode dispersion relations and the reflectivity under prism illumination. Experiment and theory are in excellent agreement for the dispersion relations and the mode propagation distances.
Our results bear some resemblance to guided modes in photonic crystal waveguides \cite{JVF97,LCH98,JFV99} and electromagnetic surface waves in drilled metal films \cite{UT1973}.

\begin{figure}
\centering\includegraphics[width=130mm]{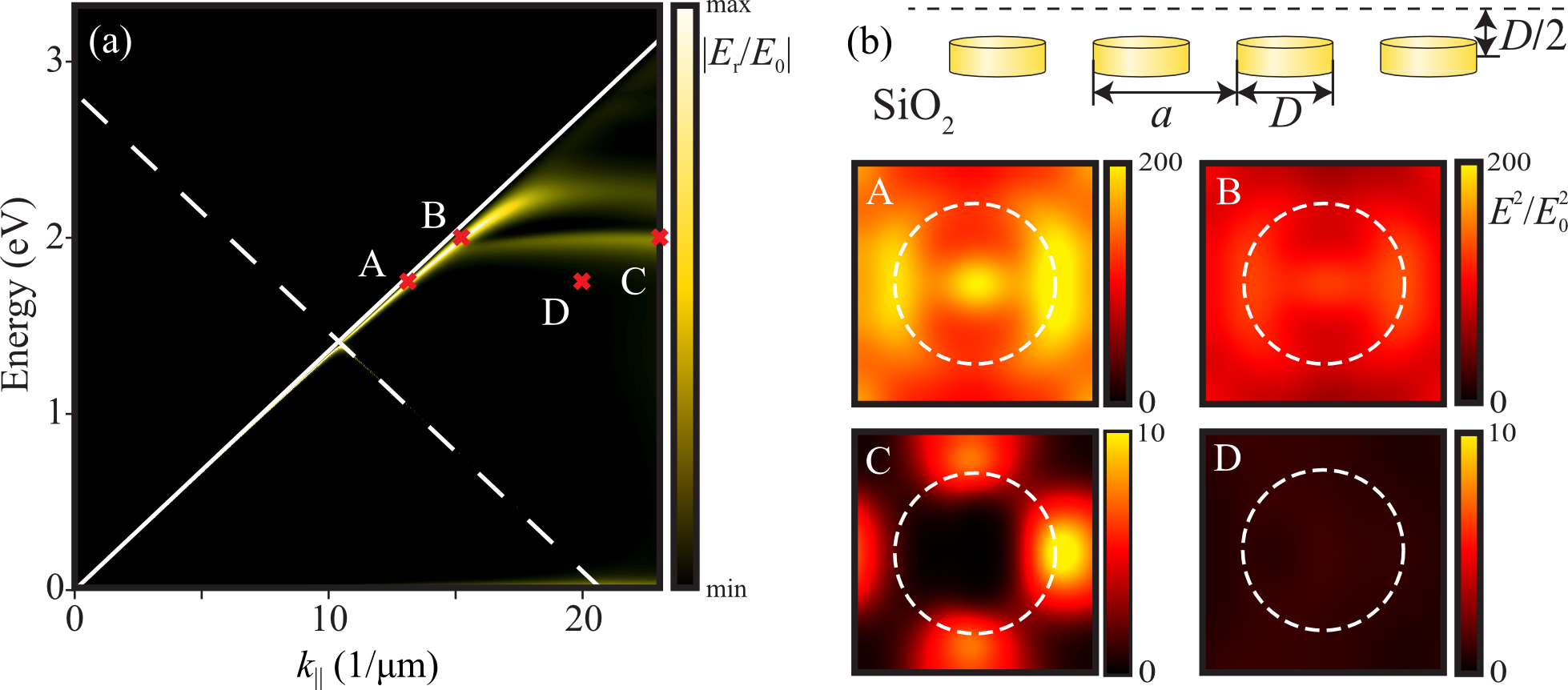}%
\caption{Simulations of confined modes in a square array of SiO$_2$-embedded gold nanodisks of $D=200\,{\rm nm}$ in diameter. {\bf(a)} Dispersion diagram showing the light energy and wave-vector dependence of the array reflection coefficient (color scale). Guided modes show up as bright features.
The light cone (solid line) and its first Bragg-reflection (dashed line) are plotted for reference. {\bf(b)} Electric-field intensity maps over a unit cell at a distance $D/2$ above the array for characteristic points within the dispersion diagram, as shown by labels A-D in both (a) and (b).
The nanoparticle contour is represented by a dashed circle. Dipolar patterns are visible when confined modes are excited (A and B), leading to large field enhancement (see color scale).
In contrast, lower field enhancement is associated with non-resonant reflection (C and D), although hotspots can localize between particle gaps (see D), leading to an almost dispersionless feature in (a).\label{fig:sym}}
\end{figure}

The array modes under study originate in the lattice resonances produced by long-range interaction between the particles close to the onset of different diffraction orders \cite{paper142}.
Under such conditions, there is constructive interference of the fields scattered by the particles, leading to array-confined modes, which are in fact guided modes that propagate along the plane of the array.
These modes show up at frequencies to the red with respect to the  light line in a frequency-wavevector diagram \cite{paper090}. In Fig.\ \ref{fig:sym}(a) we show the calculated reflection coefficient of arrays of gold nanodisks placed in a symmetric SiO$_2$ environment.
The dispersion diagram is dominated by two types of features: a prominent one arising from the noted lattice resonances; and a weaker one associated with the excitation of localized surface plasmons that are confined at the gaps between neighboring nanoparticles \cite{NOP04} (see below).
The reflection coefficient is peaked along a line close to the light cone (solid line), indicating the position of the dispersion relation of the lattice modes.
For reference, we also show in the diagram the {\it umklapp} folding of the light cone (dashed line), which crosses the light cone at the boundary of the first Brillouin zone (BZ).
While the lattice resonance within the first BZ is both too narrow and close to the light line to be resolved in the figure, it is clearly visible withing the second BZ.

The intensity of the electric field, which is represented in Fig.\ \ref{fig:sym}(b) for a unit cell at a distance $D/2$ from the middle plane of the lattice, clearly shows the collective character of the resonance.
Along the dispersion relation of the propagating mode, such at positions A and B, we have large values of the field enhancement spreading all over the plane, with relatively small spatial variations. We also represent the field at the weaker, relatively flat spectral feature below the resonance at point C.
Clearly, the latter shows strong localization at the particle gaps, which is consistent with an interpretation in terms of localized gap plasmons between neighboring nanoparticles and explains the observed vanishing group velocity.
Under non-resonant incidence conditions (i.e., away from the propagating mode and localized plasmon regions), the enhancement is featureless and weak, as shown for point D.

\begin{figure}
\centering\includegraphics[width=85mm]{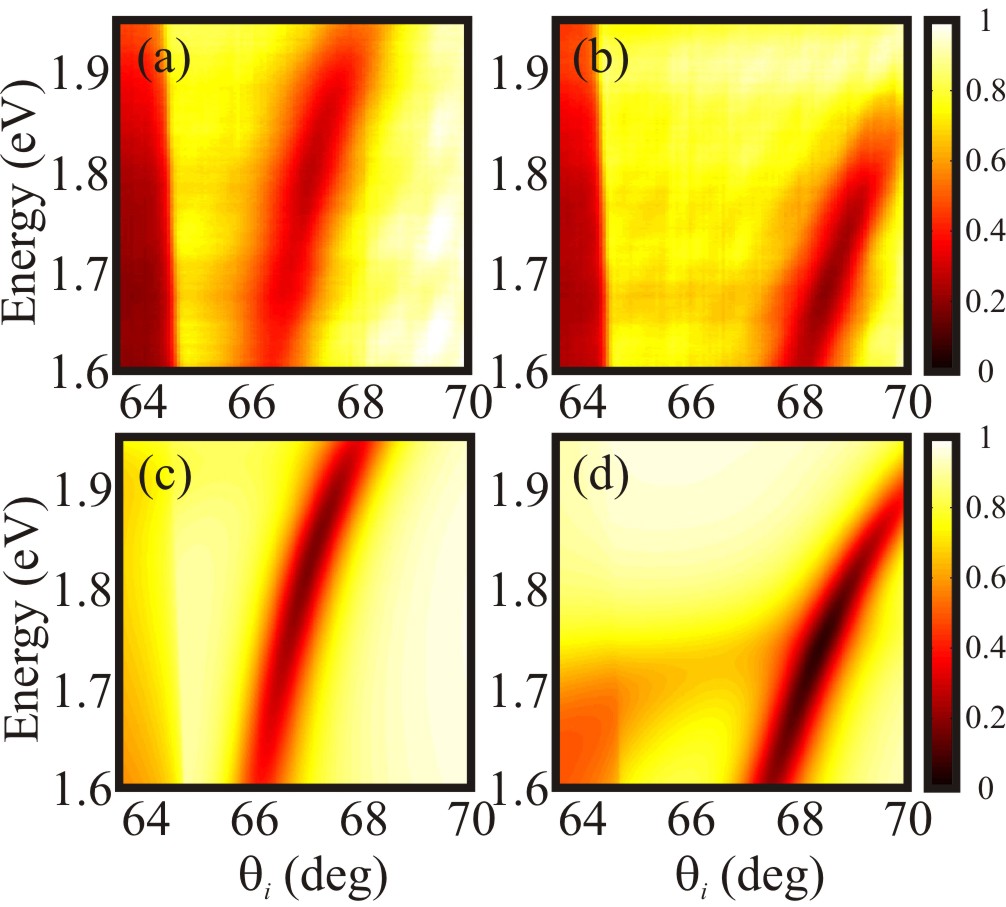}
\caption{Experimental (a,b) and theoretical (c,d) energy-angle-dependent reflectance of the combined array-prism system [Fig.\ \ref{fig:scheme}(a)] for nanodisks of diameter $D=200\,{\rm nm}$ (a,c) and $D=250\,{\rm nm}$ (b,d).
The excitation of a guided mode is revealed by a dark (low reflectance) feature at the right of the critical angle ($\theta_i\approx64\,\rm{deg}$).\label{fig:measure}}
\end{figure}

Arrays embedded in a symmetric environment thus display sharp resonances that can be easily modeled theoretically, but their fabrication and characterization is difficult compared to arrays supported on a substrate.
A high degree of asymmetry is known to strongly modify and even prevent the existence of the lattice resonances \cite{paper142,paper164}.
In order to reach a compromise between these two extreme situations, we have fabricated gold nanodisk arrays on a SiO$_2$ substrate and added a dielectric capping of the same material (\ref{fig:scheme}(a)).
This capping of thickness $d=600\,$nm not only ensures the existence of the resonant modes, but also provides protection for the nanostructure in potential device applications.
We have measured the reflectivity spectra of these samples in a rotation stage using a coupling prism of F2 glass.
When light is reflected at the prism-sample interface under total-internal reflection conditions, it produces an evanescent field that,for values of the parallel momentum matching those of the guided modes, is coupled into the latter.
This coupling transfers energy that is dissipated due to losses in the metallic nanoparticles, therefore producing a dip in the reflection spectrum, which is a signature of the excitation of a surface mode \cite{O1967}.

\begin{figure}
\centering\includegraphics[width=85mm]{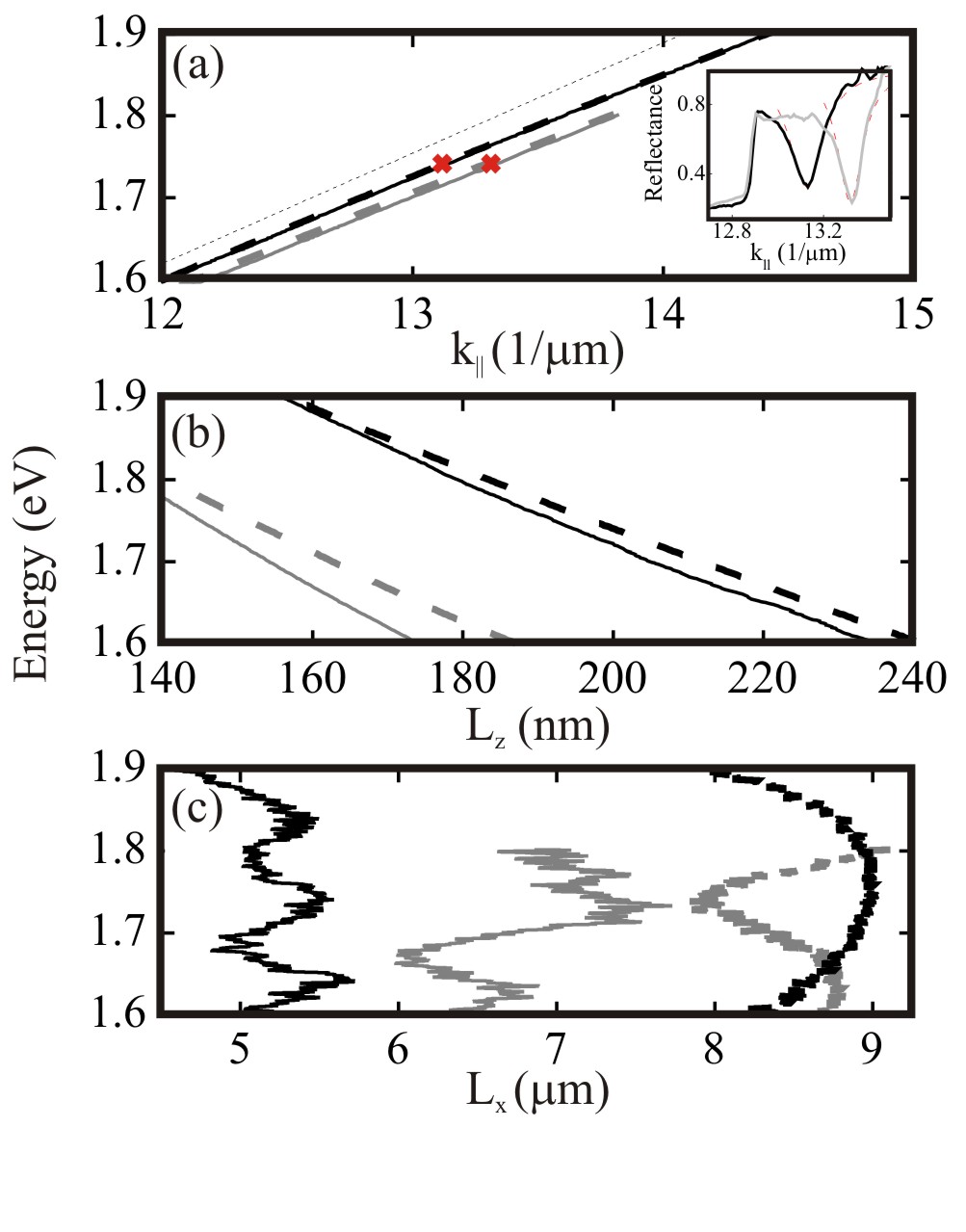}
\caption{Experiment-theory comparison of the dispersion relation {\bf(a)}, the penetration length outside the array ($L_z$) {\bf(b)}, and the propagation distance ($L_x$) {\bf(c)} for the guided modes obtained from the position and width of a Lorentzian function fitted to the data of Fig.\ \ref{fig:measure}.
Theory: dashed curves. Experiment: solid curves, obtained from constant-energy Lorentzian fits of the reflection dip in Fig. \ref{fig:measure} [the inset in (a) shows a characteristic example of the Lorentzian fit for the energy $1.74\,\rm{eV}$ signaled with crosses in the main plot].
Black (gray) curves correspond to disks of diameter $D=200\,{\rm nm}$ ($D=250\,{\rm nm}$).\label{fig:length}}
\end{figure}

The spectral dependence of this dip can be clearly resolved in Fig.\ \ref{fig:measure}, where it shows up to the right of the critical angle ($\theta_c\approx64$).
The agreement between theory [Fig.\ \ref{fig:measure}(c,d)] and experiment [Fig.\ \ref{fig:measure}(a,b)] is excellent in position, intensity, and width of the spectral features for samples with two different particle diameters [$D=200\,$nm in Fig.\ \ref{fig:measure}(a,c) and $D=250\,$nm in Fig.\ \ref{fig:measure}(b,d)].
Some discrepancies are observed in the features at the left end of the angular region under investigation, which we attribute to systematic errors in the characterization setup, which is optimized for data collection to the right of the critical angle.
As expected from the calculations of Fig.\ \ref{fig:sym} for a symmetric environment, the resonance appears close to the light line and nearly parallel to it, specially for the smallest particles. Incidentally, almost perfect coupling takes place, with the reflectance reaching values close to zero.

A more detailed characterization of the modes is presented in Fig.\ \ref{fig:length}. Panel (a) shows excellent agreement between the theoretical (dashed curves) and experimental (solid curves) dispersion relations for both particle sizes.
The degree of confinement, indicated by the separation from the light (dotted) line in Fig.\ \ref{fig:length}(a) and also by the penetration of the field intensity outside the array in Fig.\ \ref{fig:length}(b), is also in excellent agreement.
Notice that this penetration distance\cite{VGG09} is comparable to the separation of 600\,nm between the prism and the particles, but the modes are still well defined and clearly resolvable.

Finally, we have calculated the propagation distance along the plane of the array [Fig.\ \ref{fig:length}(c)] by fitting the reflectance dips to Lorentzians [see insets in Fig.\ \ref{fig:length}(a)].
Theory predicts larger propagation than what is observed in the experiment, and we attribute this discrepancy to imperfections in the shape and position of the particles.
Actually, the discrepancy between theory and experiment is less pronounced for the largest particles under consideration (gray curves), for which variations in morphology with the same degree of absolute tolerance are relatively smaller.
Propagation distances of a few microns ($\sim10\,$ surface-mode wavelengths) are obtained.

In conclusion, we have experimentally demonstrated the existence of confined propagating modes in planar arrays of plasmonic nanoparticles, and we have determined their dispersion relation and characteristic lengths. Our electromagnetic simulations are in excellent agreement with the experiments.
These results should trigger further research into the direct spatial mapping of guided modes. Our particle arrays can be potentially used to propagate optical signals.
They can also assist distant sensing, whereby the optical signal from the object or substance to be sensed propagates through an array to the position at which the detector is placed. This possibility provides a natural hybrid between plasmon-based sensors and nanoparticle-based waveguiding.

This work has been supported in part by the European Union (NMP4-2006-016881-SPANS, NMP4-SL-2008-213669-ENSEMBLE, FP7-ICT-2009-4-248909-LIMA, and FP7-ICT-2009-4-248855-N4E), the Spanish MEC (MAT2010-14885 and Consolider Nano-Light.es), and the research program of the Foundation for Fundamental Research on Matter (FOM), which is financially supported by the Netherlands Organization for Fundamental Research (NWO) and is part of an industrial partnership program between Philips and FOM.
It is also supported in part by NanoNextNL, a micro and nanotechnology consortium of the Government of the Netherlands and 130 partners. X.M.B. acknowledges a Spanish CSIC-JAE grant.

\end{document}